\documentclass[a4paper,11pt]{article}

\usepackage{enumitem}
\usepackage{float}
\usepackage[margin=2.5cm]{geometry}
\usepackage{graphicx}
\usepackage[labelfont=bf,labelsep=period]{caption}
\usepackage{times}
\usepackage{url}
\usepackage{xspace}
\usepackage{numbertabbing}

\newcommand{\ceil}[1]{\lceil #1 \rceil}

\setlist[]{topsep=2pt,partopsep=2pt,parsep=2pt,itemsep=2pt}

\floatstyle{ruled}
\newfloat{algo}{htbp}{algo}
\floatname{algo}{Algorithm}
\usepackage{cite} 

\usepackage{amsmath} 
\allowdisplaybreaks[2]          

\usepackage{amssymb} 

\usepackage{amsthm} 
\newtheoremstyle{plain-boldhead}
  {\topsep}
  {\topsep}
  {\itshape}
  {}
  {\bfseries}
  {.}
  { }
  {\thmname{#1}\thmnumber{ #2}\thmnote{ (\bfseries #3)}}
\newtheoremstyle{definition-boldhead}
  {\topsep}
  {\topsep}
  {\normalfont}
  {}
  {\bfseries}
  {.}
  { }
  {\thmname{#1}\thmnumber{ #2}\thmnote{ (\bfseries #3)}}
\theoremstyle{plain-boldhead}
\newtheorem{theorem}{Theorem}

\newtheorem{lemma}[theorem]{Lemma}

\theoremstyle{definition-boldhead}
\newtheorem{definition}{Definition}

\newtheorem{example}{Example}

\floatstyle{ruled}
\newfloat{algo}{htbp}{algo}
\floatname{algo}{Algorithm}

\renewcommand{\P}{\mathrm{P}}

\newcommand{\str}[1]{\textsc{#1}}
\newcommand{\var}[1]{\textit{#1}}
\newcommand{\op}[1]{\textsl{#1}}
\newcommand{\msg}[2]{\ensuremath{\ifempty{#2} [\str{#1}] \else [\str{#1}, {#2}] \fi}}

\newcommand{\nil}{\ensuremath{\bot}}
\newcommand{\false}{\textsc{false}\xspace}
\newcommand{\true}{\textsc{true}\xspace}

\providecommand{\keywords}[1]
{
  \small	
  \textbf{Key words.} #1
}

\newcommand{\BF}{\ensuremath{\mathbb{F}}\xspace}
\newcommand{\BK}{\ensuremath{\mathbb{K}}\xspace}

\newcommand{\BQ}{\ensuremath{\mathbb{Q}}\xspace}

\newcommand{\CA}{\ensuremath{\mathcal{A}}\xspace}
\newcommand{\CB}{\ensuremath{\mathcal{B}}\xspace}

\newcommand{\CF}{\ensuremath{\mathcal{F}}\xspace}
\newcommand{\CG}{\ensuremath{\mathcal{G}}\xspace}

\newcommand{\CK}{\ensuremath{\mathcal{K}}\xspace}

\newcommand{\CP}{\ensuremath{\mathcal{P}}\xspace}
\newcommand{\CQ}{\ensuremath{\mathcal{Q}}\xspace}

\newcommand{\CS}{\ensuremath{\mathcal{S}}\xspace}

\newcommand{\CT}{\ensuremath{\mathcal{T}}\xspace}

\providecommand{\naive}{na\"{i}ve\xspace}
\providecommand{\Naive}{Na\"{i}ve\xspace}

\def \ifempty#1{\def\temp{#1} \ifx\temp\empty }

\begin{document}

\title{\bf From Symmetric to Asymmetric Asynchronous Byzantine Consensus}

\author{Christian Cachin$^1$\\
  University of Bern\\
  \url{cachin@inf.unibe.ch}
  \and Luca Zanolini$^1$\\
   University of Bern\\
  \url{luca.zanolini@inf.unibe.ch}
}

\date{4 June 2021}

\maketitle

\footnotetext[1]{Institute of Computer Science, University of Bern,
  Neubr\"{u}ckstrasse 10, 3012 CH-Bern, Switzerland.}

\begin{abstract}\noindent
  Consensus is arguably one of the most important notions in distributed
  computing.  Among asynchronous, randomized, and signature-free
  implementations, the protocols of Most{\'{e}}faoui \emph{et al.} (PODC
  2014 and JACM 2015) represent a landmark result, which has been extended
  later and taken up in practical systems.  The protocols achieve optimal
  resilience and takes, in expectation, only a constant expected number of
  rounds of quadratic message complexity.  Randomization is provided
  through a common-coin primitive.

  In traditional consensus protocols, all involved processes adhere to a
  global, symmetric failure model, typically only defined by bounds on the
  number of faulty processes.  Motivated by applications to blockchains,
  however, more flexible trust assumptions have recently been considered.
  In particular, with \emph{asymmetric trust}, a process is free to choose
  which other processes it trusts and which ones might collude against it.

  This paper revisits the optimal asynchronous protocol of Most{\'{e}}faoui
  \emph{et al.}\ and shows how to realize it with asymmetric trust.  The
  paper starts by pointing out in detail why some versions of this protocol
  may violate liveness.  Then it proposes a fix for the protocol that does
  not affect its properties, but lets it regain the simplicity of its
  original version (PODC 2014).  At the same time, the paper shows how to
  realize randomized signature-free asynchronous Byzantine consensus with
  asymmetric quorums.  This results in an optimal consensus protocol with
  subjective, asymmetric trust and constant expected running time.  It is
  suitable for applications to blockchains, for instance.
\end{abstract}


\section{Introduction}

Consensus represents a fundamental abstraction in distributed systems. It captures the problem of reaching agreement among multiple processes on a common value, despite unreliable communication and the presence of faulty processes. Most protocols for consensus operate under the assumption that the \emph{number} of faulty processes is limited.  Moreover, all processes in the system share this common \emph{trust assumption}. Traditionally, the trust assumption has been \emph{symmetric} in this sense: all processes adhere to the global assumption about the number of faulty processes and properties of protocols are guaranteed for all correct processes, but not for the faulty ones. Since the advent of blockchains systems, however, more flexible trust models have been introduced. The Ripple~(\url{www.ripple.com}) and Stellar~(\url{www.stellar.org}) blockchains have pioneered practical models that let each process express its \emph{own set} of trusted processes and assumptions can be more flexible than bounding only the number of faulty processes.

Motivated by this desire to make trust assumptions more flexible, Cachin and Tackmann~\cite{DBLP:conf/opodis/CachinT19} introduced \emph{asymmetric Byzantine quorum systems} as a generalization of Byzantine quorum systems. Originally defined by Malkhi and Reiter~\cite{DBLP:journals/dc/MalkhiR98}, Byzantine quorum systems capture one global, but arbitrarily complex trust relation through a so-called fail-prone system.  This permits protocols in which processes can be differentiated from each other and in which not only the number of faults is bounded.  Since Byzantine quorum systems provide a widely used abstraction for realizing practial consensus protocols for distributed systems, asymmetric quorum systems open up the possibility to implement consensus with subjective trust.  However, no consensus algorithms with asymmetric trust have been formulated so far.

In this paper, we present the first asynchronous Byzantine consensus protocol with asymmetric trust.  It uses randomization, provided by an asymmetric common-coin protocol, to circumvent the impossibility of (purely) asynchronous consensus.  Our protocol takes up the randomized and signature-free implementation of consensus by Most{\'{e}}faoui \emph{et al.}~\cite{DBLP:conf/podc/MostefaouiMR14,DBLP:journals/jacm/MostefaouiMR15}. This represents a landmark result because it has been praised for its simplicity, was the first to achieve optimal complexity, that is, expected quadratic cost in the number of processes, and does not use digital signatures.  The protocol has been extended later and taken up in practical systems, such as ``Honey Badger BFT''~\cite{DBLP:conf/ccs/MillerXCSS16}.

The protocol of Most{\'{e}}faoui \emph{et al.}, however, comes in multiple versions.  The original one, published at PODC 2014~\cite{DBLP:conf/podc/MostefaouiMR14} and where it also won the best-paper award, suffers from a subtle and little-known liveness problem~\cite{TG19}: an adversary can prevent progress among the correct processes by controlling the messages between them and by sending them values in a specific order.  The subsequent version (JACM 2015)~\cite{DBLP:journals/jacm/MostefaouiMR15} resolves this issue, but requires many more communication steps and adds considerable complexity.

Our asymmetric asynchronous Byzantine consensus protocol is based on the simpler version (PODC~2014).  We first revisit this and show in detail how it is possible to violate liveness.  We propose a method that overcomes the problem, maintains the elegance of the protocol, and does not affect its appealing properties.  Based on this insight, we show how to realize asynchronous consensus with asymmetric trust, again with a protocol that maintains the simplicity of the original approach of Most{\'{e}}faoui \emph{et al.}~\cite{DBLP:conf/podc/MostefaouiMR14}.

Asymmetric quorum systems go back to the notion of asymmetric trust introduced by Damg{\aa}rd \emph{et al.}~\cite{DBLP:conf/asiacrypt/DamgardDFN07}.  Every process in the system subjectively selects its own fail-prone system.  Depending on the choice that a correct process makes about who it trusts and who not, and considering the processes that are actually faulty during an execution, two different situations may arise.  A correct process may either make a ``wrong'' trust assumption, for example, by trusting too many processes that turn out to be faulty or by tolerating too few faults; such a process is called \emph{\naive}.  Alternatively, when the correct process makes the ``right'' trust assumption, it is called \emph{wise}.  Protocols with asymmetric trust do not guarantee the same properties for \naive processes as for wise ones.

As an additional contribution, we extend our knowledge about the relation between \naive and wise processes in protocols with asymmetric trust.  We show that, under certain conditions, guarantees can only be given for a subset of the wise processes that form a so-called \emph{guild}.  The existence of a guild is necessary for a protocol execution with asymmetric trust to terminate.

The remainder of this work is structured as follows. In Section~\ref{sec:related-works} we discuss related work.  We present our system model together with preliminaries on Byzantine quorums in Section~\ref{sec:system-model}.  In Section~\ref{sec:attack} we recall the randomized consensus protocol as originally introduced by Most{\'{e}}faoui \emph{et al.}~\cite{DBLP:conf/podc/MostefaouiMR14}, discuss the liveness issue~\cite{TG19}, and show a way to prevent it. In Section~\ref{sec:asym-trust} we recall and extend the theory behind asymmetric quorums. We define and implement asymmetric strong Byzantine consensus protocol in Section~\ref{sec:asymrandom} by extending and improving on the randomized consensus algorithm by Most{\'{e}}faoui \emph{et al.}~\cite{DBLP:conf/podc/MostefaouiMR14}. Moreover, we build a common coin based on secret sharing that works in the asymmetric-trust model and it is used in our randomized protocol. Conclusions are drawn in Section~\ref{sec:conclusions}. 

\section{Related work}
\label{sec:related-works}

Most{\'{e}}faoui \emph{et al.}~\cite{DBLP:conf/podc/MostefaouiMR14} present a signature-free round-based asynchronous consensus algorithm for binary values. It achieves optimal resilience and takes $O(n^2)$ constant-sized messages. The algorithm is randomized and randomization is achieved through a common coin as defined by Rabin~\cite{DBLP:conf/focs/Rabin83}.  Their binary consensus algorithm has been taken up for constructing the HoneyBadgerBFT protocol by Miller \emph{et al.}~\cite{DBLP:conf/ccs/MillerXCSS16}, for instance.  One important contribution of Most{\'{e}}faoui \emph{et al.}~\cite{DBLP:conf/podc/MostefaouiMR14} is a new binary validated broadcast primitive with a non-deterministic termination property; it has also found applications in other protocols~\cite{DBLP:conf/nca/CrainGLR18}.

Tholoniat and Gramoli~\cite{TG19} observe a liveness issue in the protocol by Most{\'{e}}faoui \emph{et al.}~\cite{DBLP:conf/podc/MostefaouiMR14} in which an adversary is able to prevent progress among the correct processes by controlling messages between them and by sending them values in a specific order. 

In a later work, Most{\'{e}}faoui \emph{et al.}~\cite{DBLP:journals/jacm/MostefaouiMR15} present a different version of their randomized consensus algorithm that does not suffer from the liveness problem anymore. The resulting algorithm offers the same asymptotic complexity in message and time as their previous algorithm~\cite{DBLP:conf/podc/MostefaouiMR14} but it requires more communication steps. 

Flexible trust structures have recently received a lot of attention~\cite{DBLP:conf/opodis/Garcia-PerezG18, DBLP:conf/ccs/MalkhiN019, DBLP:conf/wdag/LosaGM19, Mazieres2015TheSC, DBLP:conf/opodis/CachinT19, DBLP:conf/asiacrypt/DamgardDFN07}, primarily motivated by consensus protocols for blockchains, as introduced by Ripple~(\url{www.ripple.com}) and Stellar~(\url{www.stellar.org}).  According to the general idea behind these models, processes are free to express individual, \emph{subjective} trust choices about other processes, instead of adopting a common, global view of trust. 

Damg{\aa}rd \emph{et al.}~\cite{DBLP:conf/asiacrypt/DamgardDFN07} define the basics of \emph{asymmetric trust} for secure computation protocols. This model is strictly more powerful than the standard model with symmetric trust and abandons the traditional global failure assumption in the system. Moreover, they present several variations of their asymmetric-trust model and sketch synchronous protocols for broadcast, verifiable secret sharing, and general multi-party computation.

Mazi{\`e}res~\cite{Mazieres2015TheSC} introduces a new model for consensus called \emph{federated Byzantine agreement} (FBA) and uses it to construct the \emph{Stellar consensus protocol}~\cite{DBLP:conf/sosp/LokhavaLMHBGJMM19}. In FBA, every process declares \emph{quorums slices} -- a collection of trusted sets of processes sufficient to convince the particular process of agreement. These slices are subsets of a \emph{quorum}, which is a set of processes sufficient to reach agreement.  More precisely, a quorum is defined as a set of processes that contains one slice for each member, and all quorums constitute a \emph{federated Byzantine quorum system}~(FBQS).

\emph{Byzantine quorum systems} have originally been formalized by Malkhi and Reiter~\cite{DBLP:journals/dc/MalkhiR98} and exist in several forms; they generalize the classical quorum systems aimed at tolerating crashes to algorithms with Byzantine failures. Byzantine quorum systems assume one global shared fail-prone system.

A link between FBQS and Byzantine quorums system has been built by Garc{\'{\i}}a{-}P{\'{e}}rez and Gotsman~\cite{DBLP:conf/opodis/Garcia-PerezG18}, who implement Byzantine reliable broadcast on an FBQS. They prove that a FBQS \emph{induces} a Byzantine quorum system.

\emph{Asymmetric Byzantine quorum systems} have been introduced by Cachin and Tackmann~\cite{DBLP:conf/opodis/CachinT19} and generalize Byzantine quorum systems~\cite{DBLP:journals/dc/MalkhiR98} to the model with asymmetric trust. This work also explores properties of asymmetric Byzantine quorum systems and differences to the model with symmetric trust. In particular, Cachin and Tackmann~\cite{DBLP:conf/opodis/CachinT19} distinguish between different classes of correct processes, depending on whether their failure assumptions in an execution are correct. The standard properties of protocols are guaranteed only to so-called \emph{wise} processes, i.e., those that made the ``right'' trust choices. Protocols with asymmetric quorums are shown for Byzantine consistent broadcast, reliable broadcast, and emulations of shared memory. In contrast to FBQS, asymmetric quorum systems appear to be a natural extension of symmetric quorum systems.

Recently, Losa \emph{et al.}~\cite{DBLP:conf/wdag/LosaGM19} have formulated an abstraction of the consensus mechanism in the Stellar network by introducing \emph{Personal Byzantine quorum systems} (PBQS). In contrast to the other notions of ``quorums'', their definition does not require a global intersection among quorums. This may lead to several separate \emph{consensus clusters} such that each one satisfies agreement and liveness on its own.

Another new approach for designing Byzantine fault-tolerant (BFT) consensus protocols has been introduced by Malkhi \emph{et al.}~\cite{DBLP:conf/ccs/MalkhiN019}, namely \emph{Flexible BFT}. This notion guarantees higher resilience by introducing a new \emph{alive-but-corrupt} fault type, which denotes processes that attack safety but not liveness.  Malkhi \emph{et al.}~\cite{DBLP:conf/ccs/MalkhiN019} also define \emph{flexible Byzantine quorums} that allow processes in the system to have different faults models. 

\section{System model and preliminaries}
\label{sec:system-model}

\subsection{System model}

\paragraph*{Processes.}
We consider a system of $n$ \emph{processes} $\CP = \{p_1, \dots, p_n\}$ that communicate with each other. The processes interact by exchanging messages over reliable point-to-point links, specified below.

A protocol for \CP consists of a collection of programs with instructions
for all processes.  Protocols are presented in a modular way using the
event-based notation of Cachin \emph{et al.}~\cite{DBLP:books/daglib/0025983}.

\paragraph*{Failures.}
A process that follows its protocol during an execution is called \emph{correct}.  On the other hand, a \emph{faulty} process may crash or deviate arbitrarily from its specification, e.g., when \emph{corrupted} by an adversary; such processes are also called \emph{Byzantine}.  We consider only Byzantine faults here and assume for simplicity that the faulty processes fail right at the start of an execution.

\paragraph*{Functionalities and modularity.}
A \emph{functionality} is an abstraction of a distributed computation, either used as a primitive available to the processes or defining a service that a protocol run by the processes will provide.  Functionalities may be composed in a modular way.  Every functionality in the system is specified through its \emph{interface}, containing the \emph{events} that it exposes to applications that may call it, and through a number of \emph{properties} that define its behavior. There are two kinds of events in an interface: \emph{input events} that the functionality receives from other abstractions, typically from an application that invokes its services, and \emph{output events}, through which the functionality delivers information or signals a condition.

Multiple functionalities may be composed together modularly.  In a modular protocol implementation, in particular, every process executes the program instructions of the protocol implementations for all functionalities in which it participates.

\paragraph*{Links.} 
We assume there is a low-level functionality for sending messages over point-to-point links between each pair of processes.  In a protocol, this functionality is accessed through the events of ``sending a message'' and ``receiving a message.''  Point-to-point messages are authenticated and delivered reliably among correct processes.

Moreover, we assume FIFO ordering on the reliable point-to-point links for every pair of processes (except in Section~\ref{sec:attack}).  This means that if a correct process has ``sent'' a message $m_1$ and subsequently ``sent'' a message $m_2$, then every correct process does not ``receive'' $m_2$ unless it has earlier also ``received'' $m_1$. FIFO-ordered links are actually a very common assumption. Protocols that guarantee FIFO order on top of (unordered) reliable point-to-point links are well-known and simple to implement~\cite{10.5555/HadzilacosT93,DBLP:books/daglib/0025983}.  We remark that there is only one FIFO-ordered reliable point-to-point link functionality in the model; hence, FIFO order holds among the messages exchanged by the implementations for all functionalities used by a protocol.

\paragraph*{Time and randomization.}
In this work we consider an asynchronous system, where processes have no access to any kind of physical clock, and there is no bound on processing or communication delays. The randomized consensus algorithm delegates probabilistic choices to a \emph{common coin} abstraction~\cite{DBLP:conf/focs/Rabin83}; this is a functionality that delivers the same sequence of random binary values to each process, where each binary value has the value $0$ or $1$ with probability~$\frac{1}{2}$.

\subsection{Byzantine quorum systems}
\label{sec:asym-quorum}

Let us recall Byzantine quorums as originally introduced~\cite{DBLP:journals/dc/MalkhiR98}. We refer to them as \emph{symmetric} Byzantine quorums.

\begin{definition}[Fail-prone system]\label{def:failprone}
Let \CP be a set of processes. A \emph{fail-prone system} $\CF$ is a collection of subsets of \CP, none of which is contained in another, such that some $F \in \CF$ with $F \subseteq \CP$ is called a \emph{fail-prone set} and contains all processes that may at most fail together in some execution. 

\end{definition}

\begin{definition}[Symmetric Byzantine quorum system]\label{def:quorum}
Let \CP be a set of processes and let $\CF \subseteq 2^{\CP}$ be a \emph{fail-prone system}.
  A \emph{symmetric Byzantine quorum system} for \CF is a collection of sets of
  processes $\CQ \subseteq 2^{\CP}$, where each $Q \in \CQ$ is called a
  \emph{quorum}, such that
  \begin{description}
  \item[Consistency:] \[\forall Q_1, Q_2 \in \CQ , \forall F \in \CF: \, Q_1 \cap Q_2 \not
    \subseteq F.\]
\item[Availability:] \[\forall F \in \CF: \, \exists~Q \in \CQ: \, F \cap Q = \emptyset.\]
  \end{description}
\end{definition}

For example, under the common threshold failure model, the quorums are all sets of at least \(\ceil{\frac{n+f+1}{2}}\) processes, where $f$ is the number of processes that may fail. In particular, if $n=3f+1$, quorums have $2f+1$ or more processes.

Malkhi and Reiter~\cite{DBLP:journals/dc/MalkhiR98} refer to the above definition as \emph{Byzantine dissemination quorum system}. They also define other variants of Byzantine quorum systems. 

Note that in our notion of a quorum system, one quorum can be contained in another.

We say that a set system $\CT$ \emph{dominates} another set system $\CS$ if for each $S \in \CS$ there is some $T \in \CT$ such that $S \subseteq T$. In this sense, a quorum system for $\CF$ is \emph{minimal} whenever it does not dominate any other quorum system for $\CF$.
   
\begin{definition}[$Q^3$-condition~\cite{DBLP:journals/dc/MalkhiR98,DBLP:journals/joc/HirtM00}]
\label{def:q3}
  Let \CF be a fail-prone system. We say that \CF satisfies the \emph{$Q^3$-condition}, abbreviated
  as $Q^3(\CF)$, if it holds
  \[
    \forall F_1, F_2, F_3 \in \CF: \, \CP \not\subseteq F_1 \cup F_2 \cup F_3.
  \]
\end{definition}

This is the generalization of the threshold condition $n > 3f$ for Byzantine quorum systems.
Let $\overline{\CS} = \{ \CP \setminus S | S \in \CS \}$ be the \emph{bijective complement} of a set $\CS \subseteq 2^{\CP}$.  

\begin{lemma}[Quorum system existence~\cite{DBLP:journals/dc/MalkhiR98}]\label{lem:canon}
  Let \CF be a fail-prone system. A Byzantine quorum system for \CF exists
  if and only if~$Q^3(\CF)$. In particular, if $Q^3(\CF)$ holds, then
  $\overline{\CF}$, the bijective complement of \CF, is a Byzantine quorum
  system called \emph{canonical quorum system} of $\CF$.
\end{lemma}

Note that the canonical quorum system is not always minimal.  The canonical quorum system will play a role in Section~\ref{sec:asymrandom} for implementing a common-coin functionality with asymmetric quorums. 

Given a symmetric Byzantine quorum system~\CQ, we define a \emph{kernel}~$K$ as a set of processes that overlaps with every quorum. A kernel generalizes the notion of a \emph{core set}~\cite{DBLP:conf/icdcs/JunqueiraM03}.
 
\begin{definition}[Kernel system]
  A set $K \subseteq \CP$ is a \emph{kernel} of a quorum system~\CQ whenever it holds
  \[
    \forall Q \in \CQ: \, K \cap Q \neq \emptyset.
  \]
  This can be viewed as a \emph{consistency} property.
  
  We also define the \emph{kernel system} $\CK$ of \CQ to be the set of all kernels of $\CQ$.  Given this, the \emph{minimal kernel system} is a kernel system for which every kernel~$K$ satisfies
  \[
    \forall K' \subsetneq K, \exists~Q \in \CQ : K' \cap Q = \emptyset.
  \]
\end{definition}

For example, under a threshold failure assumption where any $f$ processes
may fail, every set of $\big\lfloor\frac{n-f+1}{2}\big\rfloor$ processes is
a kernel. In particular, $n=3f+1$ if and only if every kernel has $f+1$
processes.

\begin{lemma}\label{lem:kernelinquorum}
  For every $F \in \CF$ and for every quorum $Q \in \CQ$ there exists a kernel $K \in \CK$ such that $K \subseteq Q$.
\end{lemma}
 
\begin{proof}
  Let $\CQ$ be a quorum system for \CF and let $F \in \CF$. From the consistency property of a quorum system we have that for all $ Q_1, Q_2 \in \CQ$ it holds $Q_1 \cap Q_2 \not \subseteq F$. Then, the set $K = Q_1 \setminus F \subseteq Q_1$ intersects all quorums in $\CQ$ and is a kernel of \CQ.
\end{proof}

\section{Revisiting signature-free asynchronous Byzantine consensus}
\label{sec:attack}

In 2014, Most{\'{e}}faoui \emph{et al.}~\cite{DBLP:conf/podc/MostefaouiMR14} introduced a round-based asynchronous randomized consensus algorithm for binary values. It had received considerable attention because it was the first protocol with optimal resilience, tolerating up to $f < \frac{n}{3}$ Byzantine processes, that did not use digital signatures.  Hence, this protocol needs only authenticated channels and remains secure against a computationally unbounded adversary.  Moreover, it takes $O(n^2)$ constant-sized messages in expectation and has a particularly simple structure.  This description excludes the necessary cost for implementing randomization, for which the protocol relies on an abstract common-coin primitive, as defined by Rabin~\cite{DBLP:conf/focs/Rabin83}.

This protocol, which we call the \emph{PODC-14}
version~\cite{DBLP:conf/podc/MostefaouiMR14} in the following, suffers from
a subtle and little-known problem.  It may violate liveness, as has been
explicitly mentioned by Tholoniat and Gramoli~\cite{TG19}.  The
corresponding journal publication by Most{\'{e}}faoui \emph{et
  al.}~\cite{DBLP:journals/jacm/MostefaouiMR15}, to which we refer as the
\emph{JACM-15} version, touches briefly on the issue and goes on to present
an extended protocol.  This fixes the problem, but requires also many more
communication steps and adds considerable complexity.

In this section, we revisit the PODC-14 protocol, point out in detail how it may fail, and introduce a compact way for fixing it.  We discovered this issue while extending the algorithm to asymmetric quorums.  In Section~\ref{sec:asymrandom}, we present the corresponding fixed asymmetric randomized Byzantine consensus protocol and prove it secure.  Our protocol changes the PODC-14 version in a crucial way and thereby regains the simplicity of the original approach.

Before addressing randomized consensus, we recall the key abstraction introduced in the PODC-14 paper, a protocol for broadcasting binary values.

\subsection{Binary-value broadcast}

The \emph{binary validated broadcast} primitive has been introduced in the PODC-14 version~\cite{DBLP:conf/podc/MostefaouiMR14} under the name \emph{binary-value broadcast}.\footnote{Compared to their work, we adjusted some conditions to standard terminology and chose to call the primitive ``binary \emph{validated} broadcast'' to better emphasize its aspect of validating that a delivered value was broadcast by a correct process.}
In this primitive, every process may broadcast a bit $b \in \{0,1\}$ by invoking $\op{bv-broadcast}(b)$. The broadcast primitive outputs at least one  value $b$ and possibly also both binary values through a $\op{bv-deliver}(b)$ event, according to the following notion. 

\begin{definition}[Binary validated broadcast]\label{def:bvb}
  A protocol for \emph{binary validated broadcast} satisfies the
  following properties:
\begin{description}
\item[Validity:] If at least $(f+1)$ correct processes \op{bv-broadcast} the same value~$b \in \{0,1\}$, then every correct process eventually \op{bv-delivers}~$b$.
\item[Integrity:] A correct process \op{bv-delivers} a particular value~$b$ at most once and only if $b$ has been \op{bv-broadcast} by some correct process.
\item[Agreement:] If a correct process \op{bv-delivers} some value $b$, then every correct process eventually \op{bv-delivers}~$b$.
\item[Termination:] Every correct process eventually \op{bv-delivers} some value $b$.
\end{description}
\end{definition}

The implementation given by Most{\'{e}}faoui \emph{et al.}~\cite{DBLP:conf/podc/MostefaouiMR14} works as follows. When a correct process $p_i$ invokes $\op{bv-broadcast}(b)$ for $b \in \{0,1\}$, it sends a \str{value} message containing $b$ to all processes.  Afterwards, whenever a correct process receives \str{value} messages containing $b$ from at least $f+1$ processes and has not itself sent a \str{value} message containing $b$, then it sends such message to every process.  Finally, once a correct process receives \str{value} messages containing $b$ from at least $2f+1$ processes, it delivers $b$ through $\op{bv-deliver}(b)$.  Note that a process may \op{bv-deliver} up to two values.  A formal description (in the more general asymmetric model) of this protocol appears in Algorithm~\ref{alg:abv-broadcast} in Section~\ref{sec:asymrandom}.

\subsection{Randomized consensus}

We recall the notion of \emph{randomized Byzantine consensus} here and its implementation by Most{\'{e}}faoui \emph{et al.}~\cite{DBLP:conf/podc/MostefaouiMR14}.

In a consensus primitive, every correct process proposes a value~$v$ by invoking \(\op{propose}(v)\), which typically triggers the start of the protocol among processes; it obtains as output a decided value $v$ through a \(\op{decide}(v)\) event. There are no assumptions made about the faulty processes.

Observe that when working with randomized consensus, one has to formulate the termination property probabilistically. In round-based consensus algorithms, the termination property is formulated with respect to the round number~$r$ that a process executes. It requires that the probability that a correct process decides after executing infinitely many rounds approaches~1.  

\begin{definition}[Strong Byzantine consensus]\label{def:strong}
  A protocol for asynchronous \emph{strong Byzantine consensus} satisfies:

\begin{description}
\item[Probabilistic termination:] Every correct process~$p_i$ decides with probability $1$, in the sense
  that
  \[
    \lim_{r \rightarrow + \infty}
      \P[\text{a correct process $p_i$ decides by round $r$}] = 1.
  \]
  
\item[Strong validity:] A correct process only decides a value that has been proposed by some correct process.
  
\item[Integrity:] No correct process decides twice.

\item[Agreement:] No two correct processes decide differently.

\end{description}

\end{definition}

The probabilistic termination and integrity properties together imply that every correct process decides exactly once, while the agreement property ensures that the decided values are equal. Strong validity asks that if all correct processes propose the same value~$v$, then no correct process decides a value different from~$v$. Otherwise, a correct process may only decide a value that was proposed by some correct process~\cite{DBLP:books/daglib/0025983}. In a \emph{binary} consensus protocol, as considered here, only 0 and 1 may be proposed.  In this case, strong validity is equivalent to the more commonly used property of weak validity.

The implementation of the randomized consensus algorithm Most{\'{e}}faoui \emph{et al.}~\cite{DBLP:conf/podc/MostefaouiMR14} delegates its probabilistic choices to a \emph{common coin} abstraction \cite{DBLP:conf/focs/Rabin83, DBLP:books/daglib/0025983}, a random source observable by all processes but unpredictable for an adversary. A common coin is invoked at every process by triggering a \op{release-coin} event. We say that a process \emph{release}s a coin because its value is unpredictable before the first correct process invokes the coin. The value $s \in \CB$ of the coin with tag $r$ is output through an event \op{output-coin}. 

\begin{definition}[Common coin]\label{def:cc}
  A protocol for \emph{common coin} satisfies the following properties:
\begin{description}
\item[Termination:] Every correct process eventually outputs a coin value. 
  
\item[Unpredictability:] Unless at least one correct process has released the coin, no process has any information about the coin output by a correct process.
  
\item[Matching:] With probability 1 every correct process outputs the same coin value.	

\item[No bias:] The distribution of the coin is uniform over $\mathcal{B}$.

\end{description}
\end{definition}

Common-coin primitives may be realized directly by distributed protocols or with the help of a trusted entity using secret sharing~\cite{DBLP:conf/focs/Rabin83} or threshold cryptography~\cite{DBLP:journals/joc/CachinKS05}.

\begin{algo*}
\vbox{
\small
\begin{numbertabbing}\reset

  xxxx\=xxxx\=xxxx\=xxxx\=xxxx\=xxxx\=MMMMMMMMMMMMMMMMMMM\=\kill
  \textbf{State} \label{} \\
  \> $\var{round} \gets 0$: current round \label{} \\
  \> $\var{values} \gets \{\}$: set of \op{bv-delivered} binary values for
     the round  \label{}\\
  \> $\var{aux} \gets [\{\}]^n$: stores sets of values that have
     been received in \str{aux} messages in the round  \label{}\\
 \label{} \\
 \textbf{upon event} \(\op{rbc-propose}(b)\) \textbf{do}  \label{} \\
 \> \textbf{invoke} $\op{bv-broadcast}(b)$ with tag~\var{round}  \label{}\\
 \label{} \\
 \textbf{upon} $\op{bv-deliver}(b)$ with tag $r$ \textbf{such that}
    $r = \var{round}$ \textbf{do} \label{} \\
 \> $\var{values} \gets \var{values} \cup \{b\}$  \label{}\\
 \> send message \msg{aux}{\var{round}, b} to all $p_j \in \CP$ \label{} \\
 \label{} \\
 \textbf{upon} receiving a message \msg{aux}{r, b} from $p_j$
    \textbf{such that} $r = \var{round}$ \textbf{do} \label{} \\
 \> $\var{aux}[j] \gets \var{aux}[j] \cup \{b\}$  \label{}\\
 \label{} \\
 \textbf{upon exists} $B \subseteq \var{values}$ \textbf{such that} 
    \( B \neq \{\} \) \textbf{and} \(
    |\{p_j \in \CP \,|\, B = \var{aux}[j]\}| \ge n-f \) \textbf{do}
    \label{algrbc:uponb} \\
 \> $\op{release-coin}$ with tag~\var{round}  \label{}\\
 \> \textbf{wait for} $\op{output-coin}(s)$ with tag~$\var{round}$  \label{}\\
 \> \(\var{round} \gets \var{round} + 1\)  \label{}\\
 \> \textbf{if exists} $b$ \textbf{such that} $B = \{b\}$ \textbf{then}
    \` // i.e., $|B| = 1$ \label{algrbc:ifb} \\
 \> \> \textbf{if} $b = s$ \textbf{then} \label{} \\
 \>\>\> \textbf{output} \(\op{rbc-decide(b)}\) \label{}\\
 \>\> \textbf{invoke} $\op{bv-broadcast}(b)$ with tag~\var{round}
      \` // propose $b$ for the next round  \label{algrbc:bvbit}\\
 \> \textbf{else} \label{} \\
 \> \> \textbf{invoke}  $\op{bv-broadcast}(s)$ with tag~\var{round}
      \` // propose coin value $s$ for the next round  \label{algrbc:bvcoin}\\
 \> \(\var{values} \gets [\perp]^n\)  \label{}\\
 \> $\var{aux} \gets [\{\}]^n$  \label{}\\[-5ex]
\end{numbertabbing}
}
\caption{Randomized binary consensus according to Most{\'{e}}faoui \emph{et al.}~\cite{DBLP:conf/podc/MostefaouiMR14}  (code for $p_i$).}
\label{alg:rbv-consensus}
\end{algo*}

In the remainder of this section, we recall the implementation of strong Byzantine consensus according to Most{\'{e}}faoui \emph{et al.}~\cite{DBLP:conf/podc/MostefaouiMR14} in the PODC-14 version, shown in Algorithm~\ref{alg:rbv-consensus}. A correct process \emph{proposes} a binary value $b$ by invoking $\op{rbc-propose}(b)$; the consensus abstraction \emph{decides} for $b$ through an $\op{rbc-decide}(b)$ event.

The algorithm proceeds in rounds. In each round, an instance of $\op{bv-broadcast}$ is invoked. A correct process $p_i$ executes $\op{bv-broadcast}$ and waits for a value $b$ to be \op{bv-delivered}, identified by a tag characterizing the current round. When such a bit~$b$ is received, $p_i$ adds $b$ to $\var{values}$ and broadcasts $b$ through an \str{aux} message to all processes. Whenever a process receives an \str{aux} message containing~$b$ from $p_j$, it stores $b$ in a local set $\var{aux}[j]$. Once $p_i$ has received a set $B \subseteq \var{values}$ of values such that every $b \in B$ has been delivered in \str{aux} messages from at least $n-f$ processes, then $p_i$ releases the coin for the round.  Subsequently, the process waits for the coin protocol to output a binary value~$s$ through $\op{output-coin}(s)$, tagged with the current round number.

Process $p_i$ then checks if there is a single value $b$ in $B$.  If so, and if $b=s$, then it decides for value~$b$.  The process then proceeds to the next round with proposal $b$.  If there is more than one value in $B$, then $p_i$ changes its proposal to $s$.  In any case, the process starts another round and invokes a new instance of $\op{bv-broadcast}$ with its proposal.  Note that the protocol appears to execute rounds forever.

\subsection{A liveness problem}

Tholoniat and Gramoli~\cite{TG19} mention a liveness issue with the randomized algorithm in the PODC-14 version~\cite{DBLP:conf/podc/MostefaouiMR14}, as presented in the previous section. They sketch a problem that may prevent progress by the correct processes when the messages between them are received in a specific order.  In the JACM-15 version, Most{\'{e}}faoui \emph{et al.}~\cite{DBLP:journals/jacm/MostefaouiMR15} appear to be aware of the issue and present a different, more complex consensus protocol.

We give a detailed description of the problem in Algorithm~\ref{alg:rbv-consensus}.  Recall the implementation of binary-value broadcast, which disseminates bits in \str{value} messages.  According to our model, the processes communicate by exchanging messages through an asynchronous reliable point-to-point network.  Messages may be reordered, as in the PODC-14 version.

%

\begin{figure}
\begin{center}
\includegraphics[height=6cm]{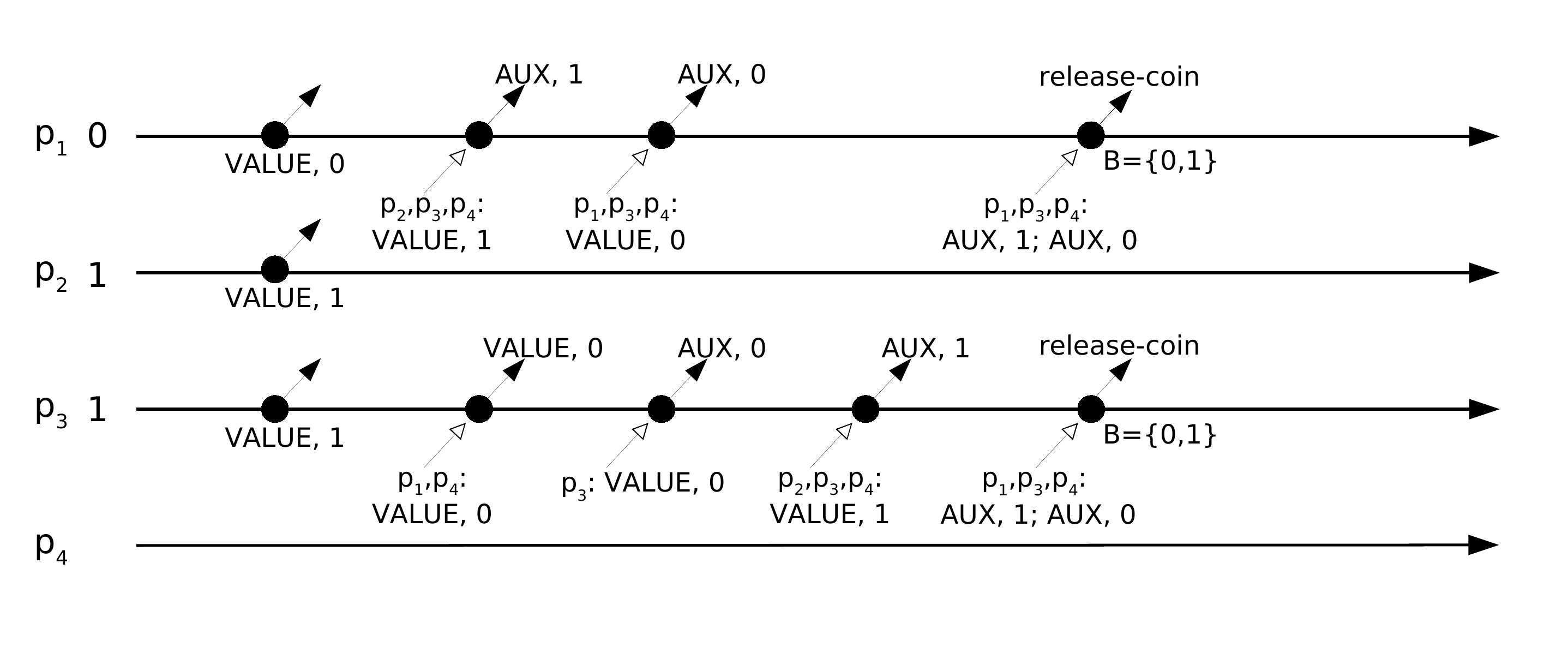}
\end{center}
\caption{The execution of Algorithm~\ref{alg:rbv-consensus}, where processes $p_1$ and $p_3$ execute line~\ref{algrbc:uponb} with $B=\{0,1\}$.}
\label{fig:attack-1}
\end{figure}

Let us consider a system with $n = 4$ processes and $f = 1$ Byzantine process.  Let $p_1, p_2$ and $p_3$ be correct processes with input values $0, 1, 1$, respectively, and let $p_4$ be a Byzantine process with control over the network. Process $p_4$ aims to cause $p_1$ and $p_3$ to release the coin with $B = \{0,1\}$, so that they subsequently propose the coin value for the next round. If messages are scheduled depending on knowledge of the round's coin value~$s$, it is possible, then, that $p_2$ releases the coin with $B = \{\neg s\}$. Subsequently, $p_2$ proposes also $\neg s$ for the next round, and this may continue forever. We now work out the details, as illustrated in Figures~\ref{fig:attack-1}--\ref{fig:attack-2}.

First, $p_4$ may cause $p_1$ to receive $2f+1$ $[\str{value},1]$ messages, from $p_2, p_3$ and $p_4$, and to $\op{bv-deliver}$~$1$ sent at the start of the round. 
Then, $p_4$ sends $[\str{value}, 0]$ to $p_3$, so that $p_3$ receives value $0$ twice (from $p_1$ and $p_4$) and also broadcasts a $[\str{value},0]$ message itself.  Process $p_4$ also sends $0$ to $p_1$, hence, $p_1$ receives $0$ from $p_3$, $p_4$, and itself and therefore $\op{bv-delivers}$ $0$.  Furthermore, $p_4$ causes $p_3$ to $\op{bv-deliver}$ $0$ by making it receive $[\str{value},0]$ messages from $p_1$, $p_4$, and itself.  Hence, $p_3$ $\op{bv-delivers}$ $0$. Finally, process $p_3$ receives three $[\str{value}, 1]$ messages (from itself, $p_2$, and $p_4$) and $\op{bv-delivers}$ also $1$.

Recall that a process may broadcast more than one $\str{aux}$ message. In particular, it broadcasts an $\str{aux}$ message containing a bit~$b$ whenever it has bv-delivered~$b$.  Thus, $p_1$ broadcasts first $[\str{aux}, 1]$ and subsequently $[\str{aux}, 0]$, whereas $p_3$ first broadcasts $[\str{aux}, 0]$ and then $[\str{aux}, 1]$.  Process $p_4$ then sends to $p_1$ and $p_3$ $\str{aux}$ messages containing $1$ and $0$. After delivering all six \str{aux} messages, both $p_1$ and $p_3$ finally obtain $B = \{0,1\}$ in line~\ref{algrbc:uponb} and see that $|B| \neq 1$ in line~\ref{algrbc:ifb}. Processes $p_1$, $p_3$ and $p_4$ invoke the common coin. 

The Byzantine process~$p_4$ may learn the coin value as soon as $p_1$ or
$p_3$ have released the common coin, according to unpredictability.  Let
$s$ be the coin output. We distinguish two cases:

\begin{description}
\item[Case $s = 0$:] Process $p_2$ receives now three $[\str{value}, 1]$ messages, from $p_3$, $p_4$ and itself, as shown in Figure~\ref{fig:attack-2}.  It $\op{bv-delivers}$~$1$ and broadcasts an $[\str{aux}, 1]$ message. Subsequently, $p_2$ delivers three $\str{aux}$ messages containing $1$, from $p_1, p_4$ and itself, but no $[\str{aux}, 0]$ message. It follows that $p_2$ obtains $B = \{1\}$ and proposes 1 for the next round in line~\ref{algrbc:bvbit}.  On the other hand, $p_1$ and $p_3$ adopt $0$ as their new proposal for the next round, according to line~\ref{algrbc:bvcoin}.  This means that no progress was made within this round. The three correct processes start the next round again with differing values, again two of them propose  one bit and the remaining one proposes the opposite.

\begin{figure}
\begin{center}
\includegraphics[height=6cm]{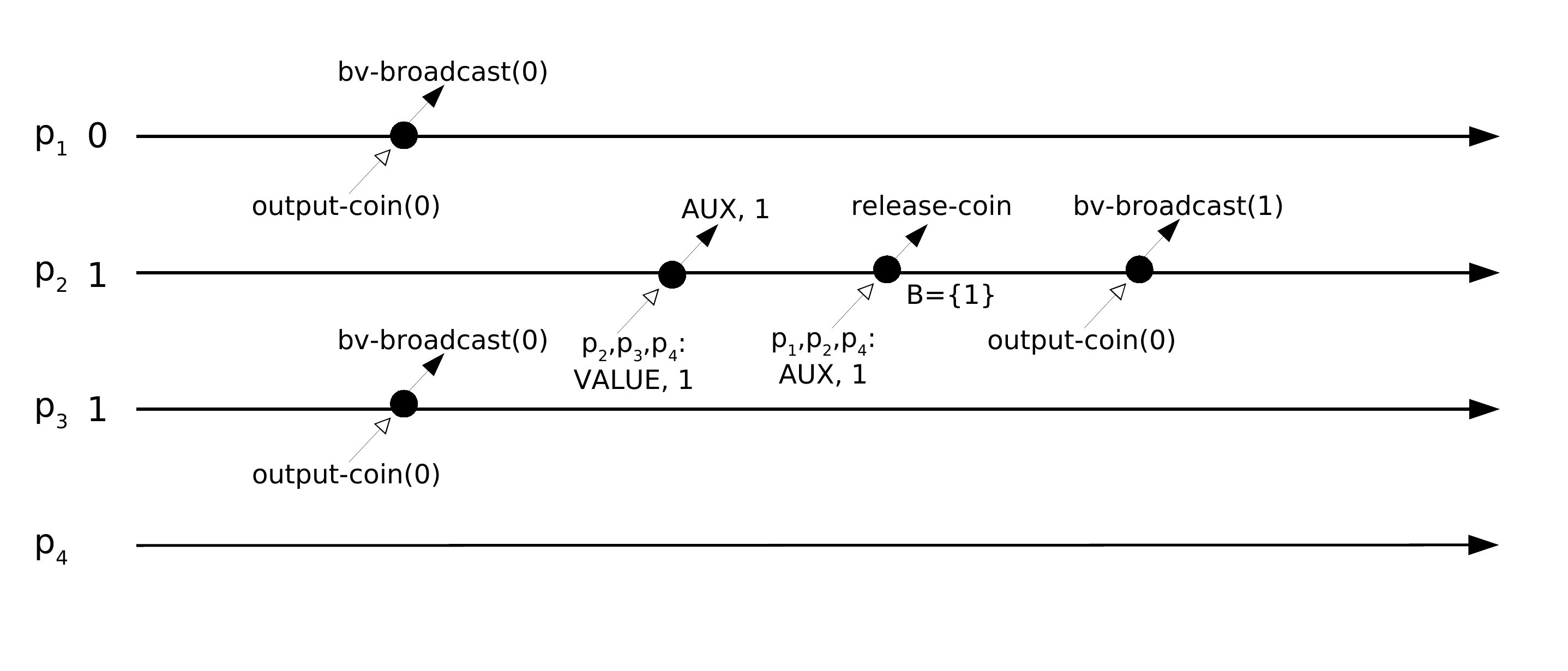}
\end{center}
\caption{Continuing the execution for the case $s=0$: Process~$p_2$ executes line~\ref{algrbc:uponb} with $B=\{1\}$.
  Processes $p_1$ and $p_3$ have already proposed the coin value $s=0$ for the next round, but $p_2$ proposes $\neg s = 1$.}
\label{fig:attack-2}
\end{figure}

\item[Case $s = 1$:] Process $p_4$ sends $[\str{value}, 0]$ to $p_2$, so that it delivers two \str{value} messages containing $0$ (from $p_1$ and $p_4$) and thus also broadcast $[\str{value}, 0]$ (this execution is not shown).  Recall that $p_3$ has already sent $[\str{value}, 0]$ before.  Thus, $p_2$ receives $n-f$ $[\str{value}, 0]$ messages, $\op{bv-delivers}$ $0$, and also broadcasts an $\str{aux}$ message containing $0$. Subsequently, $p_2$ may receive $n-f$ messages $[\str{aux}, 0]$, from $p_3$, $p_4$, and itself. It follows that $p_2$ executes line~\ref{algrbc:uponb} with $B = \{0\}$ and chooses~0 as its proposal for the next round (in line~\ref{algrbc:bvbit}). On the other hand, also here, $p_1$ and $p_3$ adopt the coin value~$s=1$ and propose~1 for the next round in line~\ref{algrbc:bvcoin}. Hence, no progress has been made in this round, as the three correct processes enter the next round with differing values.
\end{description}

The execution may continue like this forever, producing an infinite execution with no termination.

\subsection{Fixing the problem}

We show how the problem can be prevented with a conceptual insight and two small changes to the original protocol. We do this by recalling the example just presented. A formal proof is given in Section~\ref{sec:asymrandom}, using the more general model of asymmetric quorums.

We start by considering the nature of the common coin abstraction: In any full implementation, the coin is not an abstract oracle, but implemented by a concrete protocol that exchanges messages among the processes.

Observe now that in the problematic execution, the network reorders messages between correct processes. Our first change, therefore, is to assume FIFO ordering on the reliable point-to-point links. This may be implemented over authenticated links, by adding sequence numbers to messages and maintaining a buffer at the receiver~\cite{DBLP:books/daglib/0025983}.  Consider $p_2$ in the example and the messages it receives from the other correct processes, $p_1$ and $p_3$. W.l.o.g.~any protocol implementing a common coin requires an additional message exchange, where a correct process sends at least one message to every other process, say, a \str{coin} message with arbitrary content (to be specific, see Algorithm~\ref{alg:acc}, Section~\ref{sec:asymrandom}).

When $p_2$ waits for the output of the coin, it needs to receive, again w.l.o.g., a \str{coin} message from $n-f$ processes.  Since the other two correct processes ($p_1$ and $p_3$) have sent two \str{value} messages and \str{aux} messages each before releasing the coin, then $p_2$ receives these messages from at least one of them before receiving enough \str{coin} messages, according to the overlap among Byzantine quorums.

This means that $p_2$ cannot satisfy the condition in line~\ref{algrbc:uponb} with $|B| = 1$.  Thus the adversary may no longer exploit its knowledge of the coin value to prevent termination. (Most{\'{e}}faoui \emph{et al.}~\cite{DBLP:journals/jacm/MostefaouiMR15} (JACM-15) remark in retrospect about the PODC-14 version that a ``fair scheduler'' is needed.  However, this comes without any proof and thus remains open, especially because the JACM-15 version introduces a much more complex version of the protocol.)

Our second change is to allow the set $B$ to dynamically change while the coin protocol executes. In this way, process $p_2$ may find a suitable $B$ according to the received \str{aux} messages while concurrently running the coin protocol.  Eventually, $p_2$ will have output the coin \emph{and} its set $B$ will contain the same values as the sets $B$ of $p_1$ and $p_3$. Observe that this dynamicity is necessary; process $p_2$ could start to release the coin after receiving $n-f$ \str{aux} messages containing only the value $1$. However, following our example, due to the assumed FIFO order, it will receive from another correct process also an $\str{aux}$ message containing the value $0$, before the \str{coin} message. If we do not ask for the dynamicity of the set $B$, process $p_2$, after outputting the coin, will still have $|B|=1$. Most{\'{e}}faoui \emph{et al.} in the PODC-14 version (Figure~2, line~5~\cite{DBLP:conf/podc/MostefaouiMR14}) seem to rule this out. 

In Section~\ref{sec:asymrandom}, we implement these changes in a generalization of the PODC-14 version, show that the liveness problem presented in this section  no longer applies (Lemma~\ref{lem:attack}), and give a formal proof for the correctness of our protocol (Theorem~\ref{thm:arbvtheo}). The generalization works in the asymmetric-trust model, as defined in the following section.

\section{Asymmetric trust}
\label{sec:asym-trust}

In this section, we first review and extend the model of asymmetric trust, as introduced by Damg{\aa}rd \emph{et al.}~\cite{DBLP:conf/asiacrypt/DamgardDFN07} and by Cachin and Tackmann~\cite{DBLP:conf/opodis/CachinT19}. We first recall asymmetric quorums. Then we focus on a \emph{maximal guild}, which is needed for ensuring liveness and consistency in protocols, and we prove that the maximal guild is unique.  We also characterize it in relation to \emph{wise} processes, which are those correct processes whose a failure assumption turns out to be right.

In the asymmetric-trust model, every process is free to make its own trust
assumption, expressing it through a subjective fail-prone system.

\begin{definition}[Asymmetric fail-prone system]\label{def:asymfailprone}
An asymmetric fail-prone system $\BF = [\CF_1, \dots, \CF_n]$ consists of an array of fail-prone systems, where $\CF_i \subseteq 2^{\CP}$ denotes the trust assumption of $p_i$. 
\end{definition}

One often assumes that $\forall F \in \mathcal{F}_i: p_i \notin F$ for practical reasons, but this is not necessary.
For a system $\CA \subseteq 2^\CP$, let $\CA^*= \{ A' | A' \subseteq A, A \in \CA \}$ denote the collection of
all subsets of the sets in $\CA$.

\begin{definition}[Asymmetric Byzantine quorum system]\label{def:asymquorum}
  Let $\BF = [\CF_1, \dots, \CF_n]$ be an asymmetric fail-prone system. An
  \emph{asymmetric Byzantine quorum system} for \BF is an array of
  collections of sets $\BQ = [\CQ_1, \dots, \CQ_n]$, where
  $\CQ_i \subseteq 2^{\CP}$ for $i \in [1,n]$.  The set
  $\CQ_i \subseteq 2^{\CP}$ is called the \emph{quorum system of $p_i$} and
  any set $Q_i \in \CQ_i$ is called a \emph{quorum (set) for $p_i$}
  whenever the following conditions hold:
  \begin{description}
  \item[Consistency:]
    $\forall i,j \in [1,n]$
    \[
      \forall Q_i \in \CQ_i, \forall Q_j \in \CQ_j,
      \forall F_{ij} \in {\CF_i}^* \cap {\CF_j}^*: \,
      Q_i \cap Q_j \not\subseteq F_{ij}.
    \]
  \item[Availability:] 
    $\forall i \in [1,n]$
    \[
      \forall F_i \in \CF_i: \, \exists~Q_i \in \CQ_i: \, F_i \cap Q_i =
      \emptyset.
    \]
  \end{description}
  In other words, the intersection of two quorums for any two processes
  contains at least one process for which neither process assumes that it may
  fail.  Furthermore, for all fail-prone sets of every process, there exists
  a disjoint quorum for this process.
\end{definition}

The following property generalizes the $Q^3$-condition from
Definition~\ref{def:q3} to the asymmetric-trust model.

\begin{definition}[$B^3$-condition~\cite{DBLP:conf/asiacrypt/DamgardDFN07,DBLP:conf/opodis/CachinT19}]
\label{def:b3}
  Let \BF be an asymmetric fail-prone system. We say that \BF satisfies the
  \emph{$B^3$-condition}, abbreviated as $B^3(\BF)$, whenever it holds for
  all $i,j \in [1,n]$ that
  \[
    \forall F_i \in \CF_i, \forall F_j\in\CF_j,
    \forall F_{ij} \in {\CF_i}^*\cap{\CF_j}^*: \,
    \CP \not\subseteq F_i \cup F_j \cup F_{ij}. 
  \]
\end{definition}

An asymmetric fail-prone system satisfying the $B^3$-condition is
sufficient for the existence of a corresponding asymmetric
quorum system~\cite{DBLP:conf/opodis/CachinT19}.

\begin{theorem}\label{thm:asymcanon}
  An asymmetric fail-prone system \BF satisfies $B^3(\BF)$ if and only if
  there exists an asymmetric quorum system for \BF.
\end{theorem}

For implementing consensus, we also need the notion of an asymmetric kernel
system.

\begin{definition}[Asymmetric kernel system]
Let \(\BQ = [ \CQ_1, \dots, \CQ_n ]\) be an asymmetric quorum system. 
An \emph{asymmetric kernel system}~\BK is an array of
collections of sets \([ \CK_1, \dots, \CK_n ]\) such that each \(\CK_i\) is
a kernel system of \(\CQ_i\).  We call a set
$K_i \in \CK_i$ a \emph{kernel for $p_i$}.
\end{definition}

In traditional Byzantine quorum systems, under a symmetric-trust assumption, every process in the system adheres to a global fail-prone system~\CF and the set $F$ of faults or corruptions occurring in a protocol execution is in~\CF. Given this common trust assumption, properties of a protocol are guaranteed at each correct process, while they are not guaranteed for faulty ones. With asymmetric quorums, there is a distinction among correct processes with respect to $F$, namely the correct processes that consider $F$ in their trust assumption and those who do not. Given a protocol execution, the processes are classified in three different types:
\begin{description}
\item[Faulty:] A process $p_i \in F$ is \emph{faulty}.
\item[\Naive:] A correct process $p_i$ for which $F \not\in {\CF_i}^*$
  is called \emph{\naive}.
\item[Wise:] A correct process $p_i$ for which $F \in {\CF_i}^*$ is called
  \emph{wise}.
\end{description}

Recall that all processes are wise under a symmetric-trust assumption. Protocols for asymmetric quorums cannot guarantee the same properties for \naive processes as for wise ones.

A useful notion for ensuring liveness and consistency for protocols is that of a \emph{guild}. This is a set of wise processes that contains at least one quorum for each member.

\begin{definition}[Guild]
  Given a fail-prone system \BF, an asymmetric quorum system \BQ for \BF,
  and a protocol execution with faulty processes~$F$, a \emph{guild \CG for
    $F$} satisfies two properties:
  \begin{description}
  \item[Wisdom:] \CG consists of wise processes,
    \[
      \forall p_i \in \CG :\, F \in {\CF_i}^*.
    \]
  \item[Closure:] \CG contains a quorum for each of its members,
    \[
      \forall p_i \in \CG, \exists~Q_i \in \CQ_i :\, Q_i \subseteq \CG.
    \]
  \end{description}
\end{definition}

\noindent The following lemma shows that every two guilds intersect.

\begin{lemma}\label{lem:guildunique}
  In any execution with a guild $\CG$, there cannot exist two disjoint guilds.
\end{lemma}

\begin{proof}
  Let $\CP$ be a set of processes, $\CG$ be a guild and $F$ be the set of actually faulty processes. Furthermore, suppose that there is another guild $\CG'$, with $\CG \cap \CG' = \emptyset $. Let $p_i \in \CG$ and $p_j \in \CG'$ be two processes and consider a quorum $Q_i \subseteq \CG$ for $p_i$ and a quorum $Q_j \subseteq \CG'$ for $p_j$. From the definition of an asymmetric quorum system it must hold $Q_i \cap Q_j \nsubseteq F$, with $Q_i \cap Q_j \neq \emptyset$ and $F \in {\CF_i}^* \cap {\CF_j}^*$. It follows that there exists a wise process $p_k \in Q_i \cap Q_j$ with $p_k \in \CG$ and $p_k \in \CG'$. Notice also that $\CG$ and $\CG'$ both contain a quorum for $p_k$. 
\end{proof}

Observe that the union of two guilds is again a guild. It follows that every execution with a guild contains a unique \emph{maximal guild} $\CG_{\text{max}}$. Analogously to the other asymmetric notions, for a given asymmetric fail-prone system, we call the list of canonical quorum systems of all processes an \emph{asymmetric canonical quorum system}.

The following lemma shows that if a guild exists, then there cannot be a quorum for any process $p_j$ containing only faulty processes.

\begin{lemma}\label{lem:quorumonlybyz}
  Let $\CG_{\text{max}}$ be the maximal guild for a given execution and let $\mathbb{Q}$ be the canonical asymmetric quorum system. Then, there cannot be a quorum $Q_j \in \mathcal{Q}_j$ for any process $p_j$ consisting only of faulty processes.
\end{lemma}

\begin{proof}
  Given an execution with $F$ as set of faulty processes, suppose there is a guild  $\CG_{\text{max}}$. This means that for every process $p_i \in  \CG_{\text{max}}$, a quorum $Q_i \subseteq  \CG_{\text{max}}$ exists such that $Q_i \cap F = \emptyset$. It follows that for every $p_i \in  \CG_{\text{max}}$, there is a set $F_i \in \mathcal{F}_i$ such that $F \subseteq F_i$. Recall that since \BQ is a quorum system, $B^3(\BF)$ holds. From Definition~\ref{def:b3}, we have that for all $i,j \in [1,n]$, $\forall F_i \in \CF_i, \forall F_j\in\CF_j,   \forall F_{ij} \in {\CF_i}^*\cap{\CF_j}^*: \, \CP \not\subseteq F_i \cup F_j \cup F_{ij}$. 
  
  Towards a contradiction, assume that there is a process $p_j$ such that there exists a quorum $Q_j \in \mathcal{Q}_j$ for $p_j$ with $Q_j = F$. This implies that there exists $F_j \in \mathcal{F}_j$ such that $F_j = \mathcal{P} \setminus F$.
  
  Let $F_i$ be the fail-prone system of $p_i \in \CG_{\text{max}}$ such that $F \subseteq F_i$ and let $F_j = \mathcal{P} \setminus F$ as just defined. Then, $F_i \cup F_j \cup F_{ij} = \mathcal{P}$. This follows from the fact that $F_i$ contains $F$ and $F_j = \mathcal{P} \setminus F$. This contradicts the $B^3$-condition for \BF.
\end{proof}

\begin{lemma}\label{lem:quorumnaive}
  Let $\CG_{\text{max}}$ be the maximal guild for a given execution and let
  $p_i$ be any correct process. Then, every quorum for $p_i$ contains at
  least one process from the maximal guild.
\end{lemma}

\begin{proof}
  The proof naturally derives from the consistency property of an
  asymmetric quorum system.  Consider any correct process $p_i$ and one of
  its quorums, $Q_i \in \CQ_i$.  For any process
  $p_j \in \CG_{\text{max}}$, let $Q_j$ be a quorum of $p_j$ such that
  $Q_j \subseteq \CG_{\text{max}}$, which exists because $\CG_{\text{max}}$
  is a guild.  Then, the quorum consistency property implies that
  $Q_i \cap Q_j \neq \emptyset$.  Thus, $Q_i$ contains a (correct) process
  in the maximal guild.
\end{proof}

Finally, we show with an example that it is possible for a wise process to be outside the maximal guild.

\begin{example}
\label{ex:wisegmax}
Let us consider a seven-process asymmetric quorum system $\BQ_C$, defined through its fail-prone system~$\BF_C$. The notation $\Theta^n_k(\CS)$ for a set $\CS$ with $n$ elements denotes the \emph{threshold} combination operator and enumerates all subsets of $\CS$ of cardinality $k$. The diagram below shows fail-prone sets as shaded areas and the notation $\mbox{}_{k}^{n}$ in front of a fail-prone set stands for $k$ out of the $n$ processes in the set. The operator $\ast$ for two sets satisfies $\mathcal{A} \ast \mathcal{B} = \{A \cup B :~A \in \mathcal{A}, B \in \mathcal{B}\}$.\\

 \vspace*{-1ex}
\begin{minipage}[c]{0.05\linewidth}
  \vspace*{2ex}
  \center\Large$\BF_C$:
\end{minipage}
\begin{minipage}[c]{0.5\linewidth}
  \begin{eqnarray*}
    \mbox{}\\
    \CF_1 & = & \Theta^3_2(\{p_2,p_4,p_5\}) \ast \{p_6\} \ast \{p_7\}\\
    \CF_2 & = & \Theta^3_2(\{p_3,p_4,p_5\}) \ast \{p_6\} \ast \{p_7\} \\
    \CF_3 & = & \Theta^3_2(\{p_1,p_4,p_5\}) \ast \{p_6\} \ast \{p_7\}\\
    \CF_4 & = & \Theta^4_1(\{p_1,p_2,p_3,p_5\}) \ast \{p_6\} \ast \{p_7\}\\
    \CF_5 & = & \Theta^4_1(\{p_1,p_2,p_3,p_4\}) \ast \{p_6\} \ast \{p_7\}\\
    \CF_6 & = & \Theta^3_3(\{p_1,p_3,p_7\})\\
     \CF_7 & = & \Theta^3_3(\{p_3,p_4,p_5\})\\
    \mbox{}
  \end{eqnarray*}
\end{minipage}
\begin{minipage}[c]{0.5\linewidth}
  \includegraphics[height=5cm]{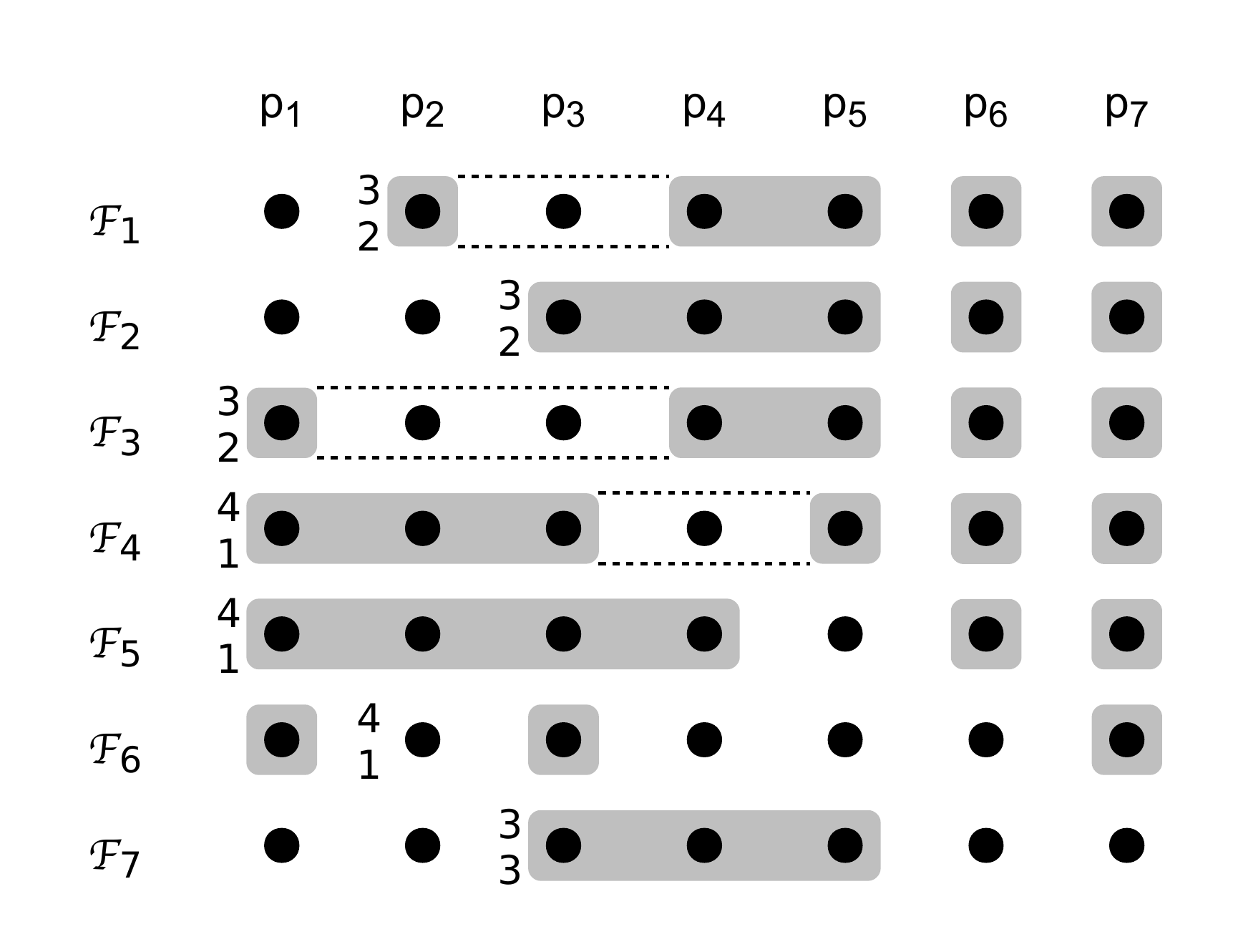}
\end{minipage}

\end{example}

One can verify that $B^3(\BF_C)$ holds; hence, let $\BQ_C$ be the
canonical quorum system of $\BF_C$. With $F = \{p_4, p_5\}$, for
instance, processes~$p_1, p_2, p_3$ and $p_7$ are wise, $p_6$ is \naive,
and the maximal guild is $\CG_{\text{max}} = \{p_1, p_2, p_3\}$.  It follows that process $p_7$ is wise but outside the guild $\CG_{\text{max}}$, because quorum $Q_7 \in \CQ_7$ contains the \naive process $p_6$.

\begin{minipage}[c]{0.1\linewidth}
  \vspace*{2ex}
  \center\Large$\BQ_C$:
\end{minipage}
\begin{minipage}[c]{0.5\linewidth}
  \begin{eqnarray*}
    \mbox{}\\
   \CQ_1 &=& \{\{p_1, p_3, p_5 \}, \{p_1, p_3, p_4 \}, \{p_1, p_2, p_3 \}\} \\
   \CQ_2 &=& \{\{p_1, p_2, p_5 \}, \{p_1, p_2, p_4 \}, \{p_1, p_2, p_3 \}\} \\
   \CQ_3  &=& \{\{p_2, p_3, p_5 \}, \{p_2, p_3, p_4 \}, \{p_1, p_2, p_3 \}\} \\
   \CQ_4  &=& \{\{p_1, p_2, p_3, p_4 \}, \{p_1, p_2, p_4, p_5 \},
               \{p_1, p_3, p_4, p_5 \}, \{p_2, p_3, p_4, p_5 \}\} \\
   \CQ_5  &=& \{\{p_1, p_2, p_3, p_5 \}, \{p_1, p_2, p_4, p_5 \},
               \{p_1, p_3, p_4, p_5 \}, \{p_2, p_3, p_4, p_5 \}\} \\
   \CQ_6 &=& \{\{p_2, p_4, p_5, p_6\}\} \\
   \CQ_7 &=& \{\{p_1, p_2, p_6, p_7 \}\} \\
    \mbox{}
  \end{eqnarray*}
\end{minipage}

\paragraph{On the importance of a guild in a protocol.}
Lemma~\ref{lem:quorumnaive} reveals an interesting result, i.e., that every quorum of every correct process contains at least a process inside the maximal guild. This means that $\CG_{\text{max}}$ is a kernel for every correct process. The maximal guild $\CG_{\text{max}}$ plays then a fundamental role in protocols with kernels by allowing correct processes (wise and \naive) to not halt during an execution and, especially, by helping wise processes outside the guild to reach termination. This means that whenever the processes in $\CG_{\text{max}}$ act, this has an influence on every correct process. 
The guild can then be intended as the mathematical formalization of ``sufficiently many wise processes such that it is possible to reach termination''. Assume, for example, to have only one wise process $p_i$ in an execution. This means that all of its quorums contain at least a \naive process. This may not be sufficient to conclude that every quorum of every \naive process contains $p_i$. However, by assuming the existence of $\CG_{\text{max}}$, there would exist at least a quorum made by only wise processes. 

A notion parallel to a guild is considered in Stellar consensus, called \emph{consensus cluster}~\cite{DBLP:conf/wdag/LosaGM19}, within which it is possible to reach consensus among correct processes. However, in contrast to our result, consensus clusters can be disjoint and an unique consensus cannot be reached among processes in disjoint consensus clusters.

\section{Asymmetric randomized Byzantine consensus}
\label{sec:asymrandom}

In this section we define asymmetric Byzantine consensus. Then we implement it by a randomized algorithm, which is based on the protocol of Most{\'{e}}faoui \emph{et al.}~\cite{DBLP:conf/podc/MostefaouiMR14} as introduced in Section~\ref{sec:attack}.  Our implementation also fixes the problem described there.

Our notion of Byzantine consensus uses strong validity in the asymmetric
model.  Furthermore, it restricts the safety properties of consensus
from all correct ones to \emph{wise} processes.  For implementing asynchronous
consensus, we use a system enriched with randomization.  In the asymmetric
model, the corresponding probabilistic termination property is guaranteed
only for wise processes.

\begin{definition}[Asymmetric strong Byzantine consensus]\label{def:asymstrong}
  A protocol for asynchronous \emph{asymmetric strong Byzantine consensus}
  satisfies:

\begin{description}
\item[Probabilistic termination:] In all executions with a guild, every wise
  process decides with probability $1$, in the sense
  that
  \[
    \lim_{r \rightarrow + \infty} (\P[\text{a wise process $p_i$
      decides by round $r$}]) = 1.
  \]
  
\item[Strong validity:] In all executions with a guild, a wise process only decides a value that has been proposed by some processes in the maximal guild.
  
\item[Integrity:] No correct process decides twice.

\item[Agreement:] No two wise processes decide differently.

\end{description}

\end{definition}

\paragraph*{Common coin.}

Our randomized consensus algorithm delegates its probabilistic choices to a \emph{common coin} abstraction \cite{DBLP:conf/focs/Rabin83, DBLP:books/daglib/0025983}. We define this in the asymmetric-trust model.

\begin{definition}[Asymmetric common coin]\label{def:acc}
  A protocol for \emph{asymmetric common coin} satisfies the
  following properties:
\begin{description}
\item[Termination:] In all executions with a guild, every process in the maximal guild eventually outputs a coin value. 
  
\item[Unpredictability:] Unless at least one correct process has released the coin, no process has any information about the coin output by a wise process.
  
\item[Matching:] In all executions with a guild, with probability $1$ every process in the maximal guild outputs the same coin value.	

\item[No bias:] The distribution of the coin is uniform over $\mathcal{B}$.

\end{description}
\end{definition}

An asymmetric common coin has an output domain~\CB.  Here we consider binary consensus and $\CB = \{0,1\}$. The \emph{termination} property guarantees that every process in the maximal guild eventually output a coin value that is ensured to be the same for each of them by the \emph{matching} property. The \emph{unpredictability} property ensures that the coin value is kept secret until a correct process releases the coin. Finally, the \emph{no bias} property specifies the probability distribution of the coin output. The bias and matching properties may be weakened using well-known methods.

\paragraph*{Implementing an asymmetric common coin.} Our implementation of asymmetric common coin relies on the scheme of Benaloh and Leichter \cite{DBLP:conf/crypto/Leichter88} and is shown in Algorithm~\ref{alg:acc}. Furthermore, following the approach started by Rabin \cite{DBLP:conf/focs/Rabin83}, we assume that coins are \emph{predistributed} by an ideal dealer using secret sharing, in a way that for every round~$r$ there is exactly one coin with value $s \in \{0,1\}$. Specifically, given an asymmetric quorum system~\BQ, the dealer creates random shares $s_{i_1},\dots,s_{i_{m-1}}$ for one random coin value $s \in \{0,1\}$ per round and for each $Q_i = \{p_{i_1}, \dots, p_{i_m}\} \in \mathcal{Q}_i$ of $\mathbb{Q}$ with $|Q_i| = m$.  Then the dealer sets $s_{i_m} = s + \sum_{j=1}^{m-1} s_{i_j} \mod 2$ and gives the shares to every process in $Q_i$. This ensures that any quorum can reconstruct the secret by computing the sum modulo $2$ of the shares. This procedure is done for each quorum in every quorum system of $\mathbb{Q}$. Furthermore, the dealer authenticates the shares, preventing Byzantine processes to send inconsistent bits to other processes in the quorum (omitted from Algorithm~\ref{alg:acc}). Correctness of this protocol follows easily.  Given a quorum of correct processes, every wise process can thus reconstruct the secret and all of them will output the same coin value. On the other hand, in every execution with a guild, Byzantine processes cannot recover the secret without receiving at least one share from a correct process by Lemma~\ref{lem:quorumonlybyz}.

\begin{algo*}[tbh]
\vbox{
\small
\begin{numbertabbing}\reset
  xxxx\=xxxx\=xxxx\=xxxx\=xxxx\=xxxx\=MMMMMMMMMMMMMMMMMMM\=\kill
  \textbf{State} \label{} \\
 \>  \(\var{coin}[k] \gets [\perp]^n\): for $k \in [1, |\CQ_i|]$, $\var{coin}[k][j]$ holds the share received from $p_j$ for quorum $Q_{i,k}$ of $p_i$ \label{}\\
   \label{} \\
  \textbf{upon event} \(\op{release-coin}\) \textbf{do}  \label{}\\
  \> \textbf{for all} \(p_j \in \CP\) \textbf{do}
  \` // send all shares to all process $p_j$ \label{} \\
  \>\> \textbf{for all} \(Q_{j,k} \in \CQ_j\) \textbf{do} \label{} \\
  \>\>\> \textbf{if} $p_i \in Q_{j,k}$ \textbf{then} \label{} \\
  \>\>\>\> send message \msg{coin}{s_i,Q_{j,k},r} to \(p_j\)
  \` // $s_i$ is share of $p_i$ for quorum $Q_{j,k}$ of $p_j$ \label{}\\
  \label{} \\
  \textbf{upon} receiving a message \msg{coin}{s_{j},Q_{i,k},r} from \(p_j\) \textbf{such that} $r = \var{round} \land Q_{i,k} \in \CQ_i \land
     p_j \in Q_{i,k}$ \textbf{do} \label{} \\
     \> \textbf{if} \(\var{coin}[k][j] = \bot\) \textbf{then}  \label{}\\
     \>\> \(\var{coin}[k][j] \gets s_{j}\) \label{}\\
  \label{} \\
  \textbf{upon exists} $k$ \textbf{such that}
     \( Q^*_{i} = \{p_j \in \CP |~\var{coin}[k][j] \neq \nil \} \in \CQ_i \) \textbf{do}
     \` // a quorum for $p_i$  \label{}\\
     \> $s \gets \sum_{p_j \in Q^*_i} \var{coin}[k][j]$ \label{}\\
  \> \textbf{output} $\op{output-coin(s)}$ \label{}
\end{numbertabbing}
}
\caption{Asymmetric common coin for round $\var{round}$ (code for~$p_i$)}
\label{alg:acc}
\end{algo*}

This implementation is expensive because the number of shares for one particular coin held by a process~$p_i$ is equal to the number of quorums in which $p_i$ is contained. In practical systems, one may also implement an asymmetric coin  ``from scratch'' according to the direction taken by Canetti and Rabin \cite{DBLP:conf/stoc/CanettiR93} or recently by Patra \emph{et al.}~\cite{DBLP:journals/dc/PatraCR14}. Alternatively, distributed cryptographic implementations appear to be possible, for example, as introduced by Cachin \emph{et al.}~\cite{DBLP:journals/joc/CachinKS05}. 

\subsection{Asymmetric binary validated broadcast}

We generalize the binary validated broadcast as introduced in Section~\ref{sec:attack} to the asymmetric-trust model. All safety properties are restricted to wise processes, and a guild is required for liveness. Recall that every process may broadcast a binary value $b \in \{0,1\}$ by invoking $\op{abv-broadcast}(b)$. The broadcast primitive outputs at least one  value $b$ and possibly also both binary values through an $\op{bv-deliver}(b)$ event, according to the following notion. 

\begin{definition}[Asymmetric binary validated broadcast]\label{def:abvb}
  A protocol for \emph{asymmetric binary validated broadcast} satisfies the
  following properties:
\begin{description}
\item[Validity:] In all executions with a guild, let $K_i$ be a kernel for a process $p_i$ in the maximal guild. If every process in $K_i$ is correct and
 has \op{abv-broadcast} the same value~$b \in \{0,1\}$, then every wise process eventually \op{abv-delivers}~$b$.
\item[Integrity:] In all executions with a guild, if a wise process \op{abv-delivers} some~$b$, then $b$ has been \op{abv-broadcast} by some
 process in the maximal guild. 
\item[Agreement:] If a wise process
  \op{abv-delivers} some value~$b$, then every wise process  eventually \op{abv-delivers}~$b$.
\item[Termination:] Every wise process
  eventually \op{abv-delivers} some value.
\end{description}
\end{definition}
 
Note that it guarantees properties only for processes that are wise or
even in the maximal guild.  Liveness properties also assume there exists a
guild.

\begin{algo*}[tbh]
\vbox{
\small
\begin{numbertabbing} \reset
  xxxx\=xxxx\=xxxx\=xxxx\=xxxx\=xxxx\=MMMMMMMMMMMMMMMMMMM\=\kill
  \textbf{State} \label{} \\
   \> \(\var{sentvalue} \gets [\str{false}]^2\): \(\var{sentvalue}[b]\) indicates whether $p_i$ has sent \msg{value}{b}  \label{}\\
  \>  \(\var{values} \gets [\emptyset]^n\): list of sets of received binary values \label{} \\
  \label{} \\
  \textbf{upon event} \(\op{abv-broadcast}(b)\) \textbf{do}  \label{}\\
  \> \(\var{sentvalue}[b] \gets \str{true}\)  \label{}\\
  \> send message \msg{value}{b} to all \(p_j \in \CP\)  \label{}\\
  \label{} \\
  \textbf{upon} receiving a message \msg{value}{b} from \(p_j\)
     \textbf{do}  \label{}\\
  \> \textbf{if} \(b \not\in \var{values}[j]\) \textbf{then}  \label{}\\
  \> \> \(\var{values}[j] \gets \var{values}[j] \cup \{b\}\)  \label{}\\
   \label{}\\
  \textbf{upon exists} \(b \in \{0,1\} \) \textbf{such that}
     \( \{p_j \in \CP |~b \in ~\var{values}[j]\} \in \CK_i \)
     \textbf{and}  \(\neg \var{sentvalue}[b]\) \textbf{do}
     \` // a kernel for $p_i$    \label{algabvb:kernel}\\
  \> \(\var{sentvalue}[b] \gets \str{true}\)  \label{}\\
  \> send message \msg{value}{b} to all $p_j \in \CP$  \label{}\\
  \label{} \\
  \textbf{upon exists} \(b \in \{0,1\} \) \textbf{such that}
     \( \{p_j \in \CP |~b \in ~\var{values}[j] \} \in \CQ_i \) \textbf{do}
     \` // a quorum for $p_i$  \label{}\\
  \> \textbf{output} \(\op{abv-deliver(b)}\) \label{}
\end{numbertabbing}
}
\caption{Asymmetric binary validated broadcast (code for~$p_i$)}
\label{alg:abv-broadcast}
\end{algo*}

Algorithm~\ref{alg:abv-broadcast} works in the same way as the binary validated broadcast by Most{\'{e}}faoui \emph{et al.}~\cite{DBLP:conf/podc/MostefaouiMR14} (Section~\ref{sec:attack}), but differs in the use of asymmetric quorums. The condition of receiving \str{value} messages containing $b$ from at least $f+1$ processes is replaced by receiving such messages from a kernel~$K_i$ for process~$p_i$.  Furthermore, a quorum $Q_i$ for $p_i$ is needed instead of $2f+1$ messages for abv-delivering a bit. 

\begin{theorem}\label{thm:abvbtheo}
  Algorithm~\ref{alg:abv-broadcast} implements asymmetric binary validated broadcast.
\end{theorem}

\begin{proof}
  To prove the \emph{validity} property, let us consider a kernel $K_i$ of correct processes for a process $p_i$ in the maximal guild $\CG_{\text{max}}$. Observe that from the consistency property of asymmetric quorum systems, there cannot be two quorums $Q_i \in \mathcal{Q}_i$ and $Q_j \in \mathcal{Q}_j$ for $p_i$ and $p_j$, respectively, such that every correct process in $Q_i$ has \(\op{abv-broadcast}\) $b$ and every correct process in $Q_j$ has \(\op{abv-broadcast}\) $b' \neq b$. Let us assume w.l.o.g.~that every quorum $Q_i$ for a process $p_i \in \CG_{\text{max}}$ contains some correct processes that have \(\op{abv-broadcast}\)~$b$. Then, the set $K$ containing only the correct processes that have \(\op{abv-broadcast}\) $b$ intersects every $Q_i$ for $p_i$ and is a kernel for $p_i$. According to the protocol, $p_i$ therefore sends \msg{value}{b} unless $\var{sentvalue}[b] = \str{true}$. However, if $\var{sentvalue}[b] = \str{true}$, $p_i$ has already sent \msg{value}{b}. Hence, every process in $\CG_{\text{max}}$ eventually sends \msg{value}{b}. All these processes together therefore are a kernel for every correct process by Lemma~\ref{lem:quorumnaive}. Let $p_c$ be a correct process; process $p_c$ therefore sends \msg{value}{b} unless $\var{sentvalue}[b] = \str{true}$. Then, every wise process receives a quorum for itself of values $b$ and \op{abv-delivers}~$b$. Observe that validity property requires a kernel for a process in the maximal guild, otherwise one cannot obtain the same guarantees. Recall Example~\ref{ex:wisegmax}, for instance: $\{p_7\}$ is a kernel for the wise process $p_7$ but not for any other wise process. If one would relax the requirement and permit that $K_i$ is a kernel for a wise process \emph{outside} the maximal guild, validity could not be ensured.

  For the \emph{integrity} property, let us assume an execution with a maximal guild $\CG_{\text{max}}$. Suppose first that only Byzantine processes \op{abv-broadcast}~$b$. Then, the set consisting of only these processes cannot be a kernel for any wise process. It follows that line~\ref{algabvb:kernel} of Algorithm~\ref{alg:abv-broadcast} cannot be satisfied. If only \naive processes \op{abv-broadcast}~$b$, by the definition of a quorum system and by the assumed existence of a maximal guild, there is at least one quorum for every process in $\CG_{\text{max}}$ that does not contain any \naive processes (e.g., as in Example~\ref{ex:wisegmax}). All \naive processes together cannot be a kernel for processes in $\CG_{\text{max}}$. Again, line~\ref{algabvb:kernel} of Algorithm~\ref{alg:abv-broadcast} cannot be satisfied. Finally, let us assume that a wise process $p_i$ outside the maximal guild \op{abv-broadcasts}~$b$. Then, $p_i$ cannot be a kernel for every wise process: it is not part of the quorums inside $\CG_{\text{max}}$. It follows that if a wise process \op{abv-delivers} some~$b$, then $b$ has been \op{abv-broadcast} by some
 processes in the maximal guild. 

  To show \emph{agreement}, suppose that a wise process~$p_i$ has \op{abv-delivered}~$b$.  Then it has obtained \msg{value}{b} messages from the processes in some quorum $Q_i \in \CQ_i$ and before from a kernel $K_i$ for itself. Each correct process in $K_i$ has sent \msg{value}{b} message to all other processes. Consider any other wise process~$p_j$.  Since $p_i$ and $p_j$ are both wise, we have $F \in {\CF_i}^*$ and $F \in {\CF_j}^*$, which implies $F \in {\CF_i}^* \cap {\CF_j}^*$. It follows that $K_i$ is also a kernel for $p_j$.  Thus, $p_j$ sends a \msg{value}{b} message to every process. This implies that all wise processes eventually receive a quorum for themselves of \msg{value}{b} messages and \op{abv-deliver}~$b$.

  For the \emph{termination} property, note that in any execution, every correct process \op{abv-broadcasts} some binary values. Termination then follows from the validity property.
\end{proof}

\subsection{Asymmetric randomized consensus}

In the following primitive, a correct process may \emph{propose} a binary value $b$ by invoking $\op{arbc-propose}(b)$; the consensus abstraction \emph{decides} for $b$ through an $\op{arbc-decide}(b)$ event.

\newcommand\est{\var{est}\xspace}
\begin{algo*}
\vbox{
\small
\begin{numbertabbing}\reset
  xxxx\=xxxx\=xxxx\=xxxx\=xxxx\=xxxx\=MMMMMMMMMMMMMMMMMMM\=\kill
  \textbf{State}  \label{}\\
  \> $\var{round} \gets 0$: current round  \label{}\\
  \> $\var{values} \gets \{\}$: set of \op{abv-delivered} binary values for
     the round  \label{}\\
  \> $\var{aux} \gets [\{\}]^n$: stores sets of values that have
     been received in \str{aux} messages in the round \label{} \\
    \> $\var{decided} \gets []^n$: stores binary values that have
     been reported as decided by other processes   \label{}\\
     \> $\var{sentdecide} \gets \false$: indicates whether $p_i$ has sent a \str{decide} message  \label{}\\
 \label{} \\
 \textbf{upon event} \(\op{arbc-propose}(b)\) \textbf{do}   \label{}\\
 \> \textbf{invoke} $\op{abv-broadcast}(b)$ with tag~\var{round} \label{} \\
 \label{} \\
 \textbf{upon} $\op{abv-deliver}(b)$ with tag $r$ \textbf{such that}
    $r = \var{round}$ \textbf{do}  \label{}\\
 \> $\var{values} \gets \var{values} \cup \{b\}$  \label{}\\
 \> send message \msg{aux}{\var{round}, b} to all $p_j \in \CP$  \label{}\\
  \label{}\\
 \textbf{upon} receiving a message \msg{aux}{r, b} from $p_j$
    \textbf{such that} $r = \var{round}$ \textbf{do}  \label{}\\
 \> $\var{aux}[j] \gets \var{aux}[j] \cup \{b\}$  \label{}\\
 \label{} \\
  \textbf{upon} receiving a message \msg{decide}{b} from $p_j$ \textbf{such that} $\var{decided}[j] = \bot$ \textbf{do}  \label{}\\
 \> $\var{decided}[j] = b$  \label{}\\
 \label{} \\
 \textbf{upon exists} \(b \neq \bot\) \textbf{such that} \(\{p_j \in \CP \,|\, \var{decided}[j] = b\} \in \CK_i\) \textbf{do}
 \` // a kernel for $p_i$  \label{}\\
 \> \textbf{if} $\neg\var{sentdecide}$ \textbf{then}  \label{}\\
 \>\> send message \msg{decide}{b} to all $p_j \in \CP$  \label{}\\
 \>\> $\var{sentdecide} \gets \true$ \label{}\\
  \label{}\\
  \textbf{upon exists} \(b \neq \bot\) \textbf{such that} \(\{p_j \in \CP \,|\, \var{decided}[j] = b\} \in \CQ_i\) \textbf{do}
 \` // a quorum for $p_i$  \label{}\\
 \> \(\op{arbc-decide}(b)\)  \label{} \\
 \> \textbf{halt} \label{}\\
 \label{} \\
 \textbf{upon exist}
 \(
   \{p_j \in \CP \,|\, \var{aux}[j] \subseteq \var{values}\} \in \CQ_i \) \textbf{do}
 \` // a quorum for $p_i$  \label{}\\
 \> $\op{release-coin}$ with tag~\var{round}  \label{}\\
 \label{} \\
 \textbf{upon event} $\op{output-coin}(s)$ with tag~$\var{round}$ \textbf{and exists}~$B \neq \{\}$ \textbf{such that} $ \forall~p_j \in Q_i, B = \var{aux}[j]$ \textbf{do}  \label{algarbc:coin} \\
 \> \(\var{round} \gets \var{round} + 1\)  \label{}\\
 \> \textbf{if exists} $b$ \textbf{such that} $|B| = 1 \land B = \{b\}$ \textbf{then} \label{} \\
 \>\> \textbf{if} $b = s \land \neg \var{sentdecide}$ \textbf{then}  \label{}\\
 \>\>\> send message \msg{decide}{b} to all $p_j \in \CP$ \label{} \\
      \>\>\> $\var{sentdecide} \gets \true$  \label{}\\
      \>\> \textbf{invoke} $\op{abv-broadcast}(b)$ with tag~\var{round}
      \` // propose $b$ for the next round \label{algarbc:proposeb} \\
 \> \textbf{else} \label{} \\
 \> \> \textbf{invoke} $\op{abv-broadcast}(s)$ with tag~\var{round}
      \` // propose coin value $s$ for the next round \label{algarbc:proposes} \\
 \> \(\var{values} \gets [\perp]^n\)  \label{}\\
 \> $\var{aux} \gets [\{\}]^n$ \label{}
\end{numbertabbing}
}
\caption{Asymmetric randomized binary consensus (code for $p_i$).}
\label{alg:arbv-consensus}
\end{algo*}

Algorithm~\ref{alg:arbv-consensus} differs from Algorithm~\ref{alg:rbv-consensus} in some aspects. Recall that both algorithms use a system where messages are authenticated and delivered reliably.  Importantly, we assume for Algorithm~\ref{alg:arbv-consensus} that all messages among correct processes are also delivered in FIFO order, even when they do not originate from the same protocol module.

Moreover, Algorithm~\ref{alg:arbv-consensus} allows the set $B$ to change while reconstructing the common coin (line~\ref{algarbc:coin}). This step is necessary in order to prevent the problem described in Section~\ref{sec:attack}. We prove this statement in Lemma~\ref{lem:attack}.

Finally, our protocol may disseminate \str{decide} messages in parallel to ensure termination. When $p_i$ receives a \str{decide} message from a kernel of processes for itself containing the same value~$b$, then it broadcasts a \str{decide} message itself containing $b$ to every processes, unless it has already done so.  Once $p_i$ receives a \str{decide} message from a quorum of processes for itself with the same value~$b$, it $\op{arbc-decides}(b)$ and halts.  This ``amplification'' step is reminiscent of Bracha's reliable broadcast protocol~\cite{DBLP:journals/iandc/Bracha87}.  Hence, the protocol does not execute rounds forever, in contrast to the original formulation of Most{\'{e}}faoui \emph{et al.}~\cite{DBLP:conf/podc/MostefaouiMR14} (Algorithm~\ref{alg:rbv-consensus}).

The following result shows that the problem described in Section~\ref{sec:attack} no longer occurs in our protocol.

\begin{lemma}\label{lem:attack}
  If a wise process $p_i$ outputs the coin with $B = \{0,1\}$, then
  every other wise process that outputs the coin has also $B=\{0,1\}$.
\end{lemma}

\begin{proof}
Let us assume that a wise process $p_i$ outputs the coin with $B = \{0,1\}$. This means that $p_i$ has received \str{coin} messages from a quorum $Q^{\str{coin}}_i$ for itself and has $B=\var{aux}[j]=\{0,1\}$ for all $p_j \in Q_i.$ Consider another wise process $p_j$ that also outputs the coin. It follows that $p_j$ has received \str{coin} messages from a quorum $Q^{\str{coin}}_j$ for itself as well. Observe that $p_i$ and $p_j$, before receiving the \str{coin} messages from every process in $Q^{\str{coin}}_i$ and $Q^{\str{coin}}_j$, respectively, receive all \str{aux} messages that the correct processes in these quorums have sent before the \str{coin} messages. This follows from the assumption of FIFO reliable point-to-point links. Quorum consistency implies that $Q^{\str{coin}}_i$ and $Q^{\str{coin}}_j$ have some correct processes in common. So, $p_i$ and $p_j$ receive some \str{aux} messages from the same correct process before they may output the coin. This means that if $p_i$ has $B=\{0,1\}$ after the $\op{output-coin}$ event, then every quorum $Q_j$ for $p_j$ will contain a process $p_k$ such that $\var{aux}[k]=\{0,1\}$ for $p_j$. Every wise process therefore must have $B=\{0,1\}$ before it can proceed. 
\end{proof}

The problem shown in Section~\ref{sec:attack} arose from messages between correct processes that were reordered in such a way that knowledge of the common coin value~$s$ was able to influence another correct process and cause it to deliver $\neg s$ alone. Lemma~\ref{lem:attack} above implies that our Algorithm~\ref{alg:arbv-consensus} prevents this because all wise processes arrive at the same set~$B$ when they output the coin.

\begin{theorem}\label{thm:arbvtheo}
  Algorithm~\ref{alg:arbv-consensus} implements asymmetric strong Byzantine consensus.
\end{theorem}

\begin{proof}
To prove the \emph{strong validity} property, assume that a wise process $p_i$ has decided a value $b$ in round $r$. This means that $p_i$ has received \msg{decide}{b} messages from a quorum $Q_i$ for itself. Furthermore, this also means that $B = \{b\}$ and $b$ is the same as the coin value in round~$r$. Then, $p_i$ has received $b$ from a quorum $Q_i$ for itself. Every process in $Q_i$ has received a \msg{AUX}{r,b} message and $b \in \var{values}$ has been \op{abv-delivered} from \op{abv}-broadcast instance. From the integrity property of \op{abv}-broadcast instance, $b$ has been \op{abv-broadcast} by a process in the maximal guild and, specifically, $\var{values}$ contains only values \op{abv-broadcast} by processes in the maximal guild. It follows that $b$ has been proposed by some processes in the maximal guild.

For the \emph{agreement} property, suppose that a wise process has received \msg{AUX}{r,b} messages from a quorum $Q_i$ for itself. Consider any other wise process $p_j$ that has received a quorum $Q_j$ for itself of \msg{AUX}{r,\bar{b}} messages. If at the end of round $r$ there is only one value in $B$, then from consistency property of quorum systems, it follows $b = \bar{b}$. Furthermore, if $b=s$ then $p_i$ and $p_j$ broadcast a \msg{decide}{b} message to every process and decide for $b$ after receiving a quorum of \msg{decide}{b} messages for themselves, otherwise they both $\op{abv-broadcast}(b)$ and they continue to $\op{abv-broadcast}(b)$ until $b=s$. If $B$ contains more than one value, then $p_i$ and $p_j$ proceed to the next round and invoke a new instance of  $\op{abv-broadcast}$ with~$s$. Therefore, at the beginning of the next round, the proposed values of all wise processes are equal. The property easily follows. 

The \emph{integrity} property is easily derived from the algorithm.

The \emph{probabilistic termination} property follows from two
observations. First, from the termination and the agreement properties of \op{abv-broadcast} it
follows that every wise process \(\op{abv-delivers}\) the same binary value
from the \(\op{abv-broadcast}\) instance and this value has been \(\op{abv-broadcast}\) by some processes in the maximal guild. Second, we show that with
probability $1$, there exists a round at the end of which all processes in
$\CG_{\text{max}}$ have the same proposal~$b$. If at the end of
round~$r$, every process in $\CG_{\text{max}}$ has proposed the coin value (line~\ref{algarbc:proposes}, Algorithm~\ref{alg:arbv-consensus}),
then all of them start the next round with the same
value. Similarly, if every process in $\CG_{\text{max}}$ has executed line~\ref{algarbc:proposeb} (Algorithm~\ref{alg:arbv-consensus}) they adopt the value $b$ and start the next round with the same value.

However, it could be the case that some wise processes in the maximal guild proposes $b$ and another one proposes the coin output~$s$. Observe that the properties of the common coin abstraction guarantee that the coin value is random and independently chosen. So, the random value $s$ is equal to the proposal value $b$ with probability~$\frac{1}{2}$. The probability that there exists a round $r'$ in which the coin equals the value $b$ proposed by all processes in $\CG_{\text{max}}$ during round $r'$ approaches $1$ when $r$ goes to infinity.

Let $r$ thus be some round in which every
process in $\CG_{\text{max}}$ \op{abv-broadcasts} the same bit~$b$; then,
none of them will ever change their proposal again. This is due to the fact
that every wise process invokes an \op{abv-broadcast} instance with the
same proposal $b$.  According to the validity and agreement properties of
asymmetric binary validated broadcast, every wise process
then delivers the same, unique value $b$. Hence, the proposal of every wise
process is set to $b$ and never changes. Finally, the
properties of common coin guarantee that with probability $1$ the processes
eventually reach a round in which the coin outputs $b$.  Therefore, with
probability $1$ every process in the maximal guild sends a \str{decide} message with value~$b$ to every process. This implies that it exists a quorum $Q_i \subseteq \CG_{\text{max}}$ for a process $p_i \in \CG_{\text{max}}$ such that every process in $Q_i$ has sent a \str{decide} message with value~$b$ to every process. Moreover, the set of processes in the maximal guild is a kernel for $p_i$ and for every other correct process $p_j$ (Lemma~\ref{lem:quorumnaive}). If a correct process $p_j$ receives a \str{decide} message with value~$b$ from a kernel for itself, it sends a \str{decide} message with value~$b$ to every process unless it has already done so. It follows that every wise process receives \str{decide} messages with the value~$b$ from a quorum for itself and \op{arbc-decides} for~$b$.
\end{proof}

\section{Conclusion}
\label{sec:conclusions}

As we show in this work, consensus protocols with asymmetric trust can be obtained by starting from existing, well-known protocols with symmetric trust.    Understanding subjective trust and implementing the corresponding protocols remains an interesting open problem, especially with respect to cryptographic constructions.

Moreover, while it has been shown that the existence of asymmetric quorums is characterized by the $B^3$-condition, it remains open how such a structure might arise spontaneously in a dynamic system, where processes join and leave without knowledge of each other.  One particular issue to consider is a Byzantine process that declares a fail-prone system with the sole aim of sabotaging the $B^3$-condition.  Achieving a complete characterization of asymmetric quorum systems and the corresponding protocols would allow to have systems with open and dynamic membership, in which the participants do not need to know each other from the start but still benefit from strong consistency guarantees.

\section*{Acknowledgments}

The authors thank Orestis Alpos, Vincent Gramoli, Giorgia Azzurra Marson, Achour Most{\'{e}}faoui, and anonymous reviewers for interesting discussions and helpful feedback.

This work has been funded by the Swiss National Science Foundation (SNSF)
under grant agreement Nr\@.~200021\_188443 (Advanced Consensus Protocols).

\bibliography{references, dblpbibtex}
\bibliographystyle{ieeesort}

\end{document}